\documentclass[11pt]{article}

\usepackage[margin=2.5cm]{geometry}
\usepackage{amsthm}
\usepackage{amsmath}
\usepackage{amsfonts}
\usepackage{graphicx}
\usepackage{enumerate}

\theoremstyle{definition}
\newtheorem{dfn}{Definition}[section]
\theoremstyle{remark}
\newtheorem{note}{Note}[section]
\theoremstyle{plain}

\newtheorem{ass}{Assumption}[section]

\newcommand{\cals}{\mathcal{S}}

\newcommand{\tp}{\otimes}

\newcommand{\R}{\textbf{R}}

\newcommand{\N}{\textbf{N}}
\newcommand{\C}{\textbf{C}}

\begin{document}
\title{Convolution product construction of interactions in probabilistic physical models}

\author{H. M. Ratsimbarison\\[7pt]
Institute of Astrophyics and High Energy Physics of Madagascar, \\
 Q 208, Faculty of Science Building,\\
  Antananarivo University}

\maketitle

\begin{abstract} This paper aims to give a probabilistic construction of interactions which may be relevant for building physical theories such as interacting quantum field fheories. We start with the path integral definition of partition function in quantum field theory which recall us the probabilistic nature of this physical theory. From a Gaussian law considered as free theory, an interacting theory is constructed by nontrivial convolution product between the free theory and an interacting term which is also a probability law. The resulting theory, again a probability law,  exhibits two proprieties already present in nowadays theories of interactions such as Gauge theory : the interaction term does not depend on the free term, and two different free theories can be implemented with the same interaction. 
\end{abstract}

\maketitle

\section{Introduction}

In this paper, our main motivation is to suggest a probabilistic construction of interactions in Euclidean Quantum Field Theory (QFT). QFT is a physical theory which combines field theory, already used in Classical Mechanics to describe Electromagnetism, and quantum mechanical principles, believed to govern the behavior of microscopic systems. The most computable approach of QFT is the path integral formalism where free Euclidean QFT is assumed to be a formal Gaussian probability law on a space of fields and interacting QFT extends the formal Gaussian law by adding an additional term, the interacting term, to the quadratic term of the free QFT measure.

However, in this setting, interacting QFTs suffer from divergence problems because most of relevant quantities calculated in are divergent. Although there exists a physical theory, the renormalization theory, enable to solve divergence problems, to introduce interacting term as additional term of the free term is unjustified from a probabilistic viewpoint because the resulting theory is not necessary a probability law. This paper indicates a construction of interacting theories which are --\emph{a priori}-- probability laws.
From a probability law considered as free theory, an interacting theory is constructed by convolution product between the free theory and an interacting term where this latter defines also a probability law. In this case, the interacting term does not depend on the free term and two different free theory can be implemented with the same interaction as in usual theories of interaction such as Gauge theory. When the free and interacting terms are Gaussian laws, we work out some natural conditions on the convolution product and use the exponential map to provide a general example. A calculation of the two-point function of our theory exhibits some analogous proprieties already present in the usual path integral formalism. The direct use of Gaussian measures, bypassing Lebesgue measures, allows to generalize the present construction for infinite dimensional spaces equipped with Gaussian measures such as often encountered in QFT \cite{jgaj81}.

\section{Partition functions in QFT}
 
Partition functions are the main object of the path integral approach in QFT. It allows to compute correlation fucntions which are then used to derive the S-matrix of a physical process described by an interacting QFT. For such QFTs, fundamental interactions between matters are usually explained by Gauge Theory. Its main feature is that a free Lagrangian is not invariant under some local transformations on matter fields unless one introduces a supplementary term containing a new 'field', the \emph{gauge potential}, which mediates interaction between matter fields; it is the minimal coupling procedure. In addition, one must also construct a free Lagrangian term for the gauge potential.
When one works within the path integral formalism, one defines formally the Euclidean partition function Z as
\begin{eqnarray}
 Z &:=& \int_{Fields} D\phi \int_{GP} DA \; e^{-S_{m,free}(\phi ) - S_{int}(\phi,A) }e^{-S_{g,free}(A) - S_{g,self-int}(A)},
\end{eqnarray}
with the normalization condition fixed by free theory
\begin{eqnarray}
	Z_{free} := \int_{Fields} D\phi\; e^{-S_{m,free}(\phi)} = 1,
	\label{norcon}
\end{eqnarray}
where:
\begin{itemize}
	\item S$_{m,free}$ and S$_{g,free}$ are respectively free actions of the matter field $\phi\in$Fields and the gauge potential A$\in$GP. Usually, free actions are nondegenerate positive sequilinear forms for matter fields but they are initially degenerate for gauge potentials. However, final free gauge actions, which are nondegenerate, are obtained from initial ones by adding a gauge fixing term.  
	\item S$_{int}$($\phi$,A) is the interacting term which describes fundamental interactions between matters. In the minimal coupling procedure, it is of the form
\begin{eqnarray}
	S_{int}(\phi,A) = B(\phi,\Sigma(A)\phi) \quad \phi\in Fields,\: A\in GP,
\end{eqnarray}
where B is a sesquilinear form on Fields and $\Sigma$(A) an hermitian operator on Fields for any gauge potential A. The hermiticity of $\Sigma$(A) is equivalent to the unitarity of gauge transformations on Fields.
\item S$_{g,self-int}$(A) is a self-interaction term of non-abelian gauge potentials which are present when describing some fundamental interactions such as strong interaction.
\end{itemize}
\begin{note} The above partition function is a formal object because first, the two measures D$\phi$ and DA each on infinite dimensional spaces are formal and second, it is divergent when one tries to evaluate (some part of) it.  However, after some nontrivial procedure on Z called \emph{renormalization} \cite{pdel96}, one obtains a finite quantity Z$_{ren}$ and for the probability convenience of QFT, one may normalize it. According to the normalization condition (\ref{norcon}), one may try to define a QFT as a probability law on the space of fields (after performing the integration on GP). In order to manipulate well-defined functional integrals, our strategy is to consider only functional integrals constructed from Gaussian measures.
\end{note}

\section{Sequence construction of interaction}

As seen in the first section, one may assume the existence of a Gaussian measure $\mu_{free}$ on the space of fields, and usual interacting QFTs are obtained by namely adding a supplementary (interacting) term to the free action. However, this last step conducts to divergence problems. To obtain an intertacting theory which is again a probability law, our idea is to introduce the interacting term by means of convolution product as is done in some constructions in probability theory when one deals with sequences of dependent random variables.

\subsection{Interacting sequences}

In probability theory, theorems on the weak convergence to a normal law, such as the Lindeberg-Feller theorem \cite{ribap01}, works essentially for sequences of independent random variables. More precisely, one considers a sequence of independent random variables (X$_n$)$_{n\in\N}$ and its partial sum process ($\displaystyle\sum_{i=0}^nX_i$)$_{n\in\N}$; under some additional conditions on the mean and the variance of (X$_n$)$_{n\in\N}$, the partial sum process ($\displaystyle\sum_{i=0}^nX_i$)$_{n\in\N}$ converges weakly to a normal random variable. These conditions on the mean and variance of the sequence are not so important in the sense that they do not depend on the values of these two quantities. Roughly speaking, the partial sum process of a sequence of independent random variables is inclined to follow a normal law.
\\[10pt]
On the other hand, free physical systems such as free QFTs are often described by a quadratic action, i.e. by normal laws in the path integral formalism. Therefore, one may suggest:
\begin{ass} A free physical system can be represented by the partial sum process of an independent random variables sequence. More generally, an interacting physical system can be represented by sequence of dependent random variables. 
\end{ass}
It is well-known that the probability law of a sum of independent (not necessarily equally distributed) random variables is given by the convolution product of random variable's laws. One deduces from the above explanation that a sequence of convolutions of probability laws converges weakly to a normal law when its mean and variance satisfy some technical conditions. 

Now, we will show that the probability law of an interacting sequence can also be obtained by a convolution product of its free probability law. In order to introduce ourself on the subject, it suffices to consider first discrete probability laws. 

\subsubsection{Pointwise product construction.}
For discrete probability laws, this method consists to introduce interacting term by pointwise product with the free probability. For probabilities having densities, it amounts to pointwise product the interacting term with the free probability density. Therefore, from a discrete probability law p$_{free}$ representing a free theory, we define a new probability law p by:
\begin{eqnarray}
	p = p_{free}.p_{int},
\end{eqnarray}
where . is the pointwise product of real-valued functions, and p$_{int}$ is a real function such that:
\begin{eqnarray}
	0 \leq p_{free}.p_{int} \leq 1 \quad \textrm{and} \quad
	\sum_{k\in\N} p_{free}(k)p_{int}(k) = 1. 
\end{eqnarray}
Clearly, the construction of the interacting term p$_{int}$ amounts to find a random variable with law p$_{free}$ and mean one.
\begin{note}
In the pointwise product construction, even if different interacting terms can be associated to a given free probability, their constructions depend implicitly on the free term. Moreover, it is not difficult to show that two different free probability laws cannot have the same interacting term. These two proprieties are not convenient for Particle Physics where interactions are constructed independently of the free term and different particles may have the same interaction as suggested by Gauge Theory \cite{pdel96}. 
\end{note}
One obtains analogous results for probabilities having densities, so let us move on to the next construction which would develop more appropriate proprieties.

\subsubsection{Convolution product construction.}
Another way to introduce interacting term is to multiplicate this latter with the free term by means of a convolution product. For probabilities having densities, it amounts to consider the convolution of the free probability density with the interacting term. From a discrete probability law p$_{free}$, we define a probability law \^p of an interacting sequence by:
\begin{eqnarray}
	\hat{p} = p_{free}*\hat{p}_{int},
\end{eqnarray}
where for f,g$\in$ Map($\N,\C$), the associative convolution product is defined by: 
\begin{eqnarray}
	f*g := m_{\C}\circ (f\tp g)\circ\Delta^+ ,\quad \Delta^+(k) := \sum_{\substack{a+b = k\\a,b\in \N}}a\oplus b, \: k\in \N \quad \text{(m$_{\C}$ is the multiplication on $\C$)}
	\label{stacovpr}
\end{eqnarray}
and analogous conditions to those of the pointwise construction for the real-valued function p$_{int}$, i.e. 
\begin{eqnarray}
	0 \leq p_{free}*\hat{p}_{int} \leq 1 \quad \textrm{and} \quad \sum_{k\in \N} p_{free}*\hat{p}_{int}(k) = 1.
	\label{codcovpr}
\end{eqnarray}
\begin{note} The interacting term \^p$_{int}$ is necessarily a probability law when one uses the discrete convolution product (\ref{stacovpr}). Moreover, it does not depend on the free probability law and two different free probability laws can have the same interacting term. Such proprieties are present in some constructions of interaction such as Gauge theory in Particle physics.
\end{note}
Analogous results are obtained for probabilities having densities when one uses the standard convolution product on the space L$^1$(\textbf{R}) of integrable functions defined on $\R$. It is then promising to extract more features of the above construction for measurable vector spaces.

\section{Convolution product construction on finite dimensional vector spaces}

After these discussions concerning mainly discrete laws, it is natural to consider the usual generalization of the convolution product of measures defined on a measurable vector space. In this case, the (usual) convolution product of two measures is simply given by the pushforward of their product measure under the addition map of the vector space.
\begin{dfn} Let V be a measurable vector space, $\mu_1$, $\mu_2$ two measures on V, then the \emph{convolution product} of $\mu_1$ by $\mu_2$ is the measure $\mu_1*\mu_2$ on V defined by:
\begin{eqnarray*}
  \int_V d(\mu_1*\mu_2)(x)\,f(x) &:=& \int_V d\mu_1(x)\int_V d\mu_2(y)\,f(x+y)  \quad \forall \text{ f an integrable map on V},\\
   &=& \int_V d\mu_1(x)\int_V d\mu_2(y)\,f(x),\\
	\textrm{ or } \; \mu_1*\mu_2 &:=& (\mu_1\times \mu_2)\circ C(\Sigma), 
\end{eqnarray*}
where $\mu_1\times \mu_2$ is the product measure of $\mu_1$ by $\mu_2$ defined on V$\oplus$V, and
\begin{eqnarray*}
\begin{aligned}
  \Sigma: V\oplus V &\rightarrow V\\
  x\oplus y &\mapsto x + y
\end{aligned}
\quad , \quad
\begin{aligned}
  C(\Sigma): C(V) &\rightarrow C(V\oplus V)\\
     f &\mapsto f\circ \Sigma.
\end{aligned}
\end{eqnarray*}
\label{defconv}
\end{dfn}
However, it is not difficult to show that the convolution product of two Gaussian measures on $\R$ with variances $\sigma^2$ and $\sigma'^2$ is again a Gaussian measure with variance $\sigma^2$ + $\sigma'^2$. In other words, the usual convolution product is not convenient for the construction of an interacting (i.e. non Gaussian) probability law from Gaussian laws on a measurable vector space. Moreover, in a standard interacting QFT, the partition function is given by an iterated functional integral over two different domains. Therefore, we are interested in the construction of an interacting measure from two (Gaussian) measures defined respectively on two different vector spaces. For two measures $\mu_F$ and $\mu_P$ defined on two measurable vector spaces F and P respectively, we will use the following definition:
\begin{eqnarray*}
	\int d(\mu_F*_{\zeta}\mu_P)(u)\,f(u) &:=& \int d\mu_F(u)\int d\mu_P(A)\,f(\zeta(A)^{-1}u)  \quad \forall \text{ f an integrable map on F},\\
	 &=& \int d\mu_P(A)\int d\mu_F(\zeta(A)u)\,f(u),\\
	\textrm{or } \mu_1*\mu_2 &=& (\mu_1\times \mu_2)\circ C(\Theta),
\end{eqnarray*}
where $\zeta$ is a map from P to Aut(F) and 
\begin{eqnarray*}
	\begin{aligned}
  \Theta: F\oplus P &\rightarrow F,\\
  u\oplus A &\mapsto \zeta(A)^{-1}u.
\end{aligned}
\end{eqnarray*}
Of course, one may obtain many possible convolution products following possible maps $\zeta$.

The next subsection will select generalized convolution products which may lead to good physical interpretations.

\subsection{Interaction for probabilities on finite dimensional vector spaces}

Here, we will select convolution products, by choosing the map $\zeta$, which well-behaved when we are dealing with Gaussian measures on finite dimensional vector spaces.

Let F,P be two finite dimensional complex vector spaces representing a space of 'matter fields' and a space of 'gauge potentials' respectively\footnote{However, one can recover the self-interaction case by considering F = P.}. We agree ourselves to consider positive definite sesquilinear forms B$_m$ and B$_g$ on F and P respectively as free actions which then defines Gaussian measures on these spaces.

In the minimal coupling procedure, the interacting term is of the form
\begin{eqnarray}
	S_{int}(u,A) = B(u,\Sigma(A)u) \quad u\in F,\: A\in P,
\end{eqnarray}
where B is a bilinear form on F and $\Sigma$(A) an hermitian operator on F for any gauge potential A. The hermiticity of $\Sigma$(A) is equivalent to the unitarity of gauge transformations on F.

On the other hand, our construction of interaction consists to define the partition function and the probability law of an interacting physical theory on F by means of the following convolution product:
\begin{eqnarray}
 && Z_{m*_{\zeta}g} := N(\zeta)\int_{F} du \int_{P} dA \; e^{-\frac{1}{2}B_m(\zeta(A)u,\zeta(A)u)}e^{-\frac{1}{2}B_g(A,A)},\\
	\label{scpart1}
	&\text{and}&  d\mu_{m*_{\sigma}g}(u) := N(\zeta)\,du\int_PdA\; e^{-\frac{1}{2}B_m(\zeta(A)u,\zeta(A)u)}e^{-\frac{1}{2}B_g(A,A)},\\
	&& \mu_{m*_{\zeta}g} := (\mu_m\times\mu_g)\circ C(\Theta),\quad   \Theta: u\oplus A \mapsto \zeta(A)^{-1}u,
	\label{measdef}
\end{eqnarray}
where:
\begin{enumerate}
  \item measures du and dA are respectively Lebesgue measures on F and P such that
\begin{eqnarray*}
	\int_{F} du \; e^{-\frac{1}{2}B_m(u,u)} := \int_Fd\mu_m(u) = \int_{P} dA \; e^{-\frac{1}{2}B_g(A,A)} := \int_P d\mu_g(A) = 1.
\end{eqnarray*}
It is important to note that the definition (\ref{measdef}) of $\mu_{m*_{\zeta}g}$ uses directly Gaussian measures $\mu_m$ and $\mu_g$ and then may admit a suitable generalization in infinite dimensional vector spaces equipped with Gaussian measures such as the dual Schwartz space $\cals^*$($\R^4$) \cite{jbas04}.
	\item for all A$\in$P, $\zeta$(A)$\in$ End(F). The map $\zeta$ characterizes the nature of the convolution product, in other words, those of the interaction;
	\item for all A$\in$P, B$_{m,\zeta(A)}$ := B$_m$($\zeta(A)\cdot, \zeta(A)\cdot$) is positive definite and sesquilinear. This is equivalent to consider invertible maps $\zeta$(A) for all A$\in$P. The main reason for this condition is to facilitate the normalization of Z$_{m*_{\zeta}g}$. Indeed, when we perform only the integration on F, we obtain:
\begin{eqnarray*}
	Z_{m*_{\zeta}g} = N(\zeta)\int_{P} dA \;det(\zeta(A)^*\zeta(A))^{-1/2} e^{-\frac{1}{2}B_g(A,A)},
\end{eqnarray*}
where $\zeta$(A)* is the hermitian conjugate of $\zeta$(A) with rapport to B$_m$.
	\item the determinant det($\zeta(A)^*\zeta(A)$) =: N($\zeta$)$^2$, N($\zeta$)$\geq$0, does not depend on A. With this supplementary condition, the integration on the r-dimensional vector space P is easily achieved and one obtains a normalized partition function:
\begin{eqnarray}
	Z_{m*_{\zeta}g} = N(\zeta)det(\zeta(A)^*\zeta(A))^{-1/2} = 1.
\end{eqnarray}
\end{enumerate}
The third condition is equivalent to the fact that $\zeta(A)^*\zeta(A)$ is a positive definite operator and there is a converse propriety \cite{npla98} which says that every positive definite operator is of this form. Henceforth, one can formulate an equivalent definition of the partition function given by:
\begin{eqnarray}
 Z_{m*_{\zeta}g} := N(\Xi)\int_{F} du \int_{P} dA\; e^{-\frac{1}{2}B_m(u,\Xi(A)u)}e^{-\frac{1}{2}B_g(A,A)},
	\label{scpart2}
\end{eqnarray}
where:
\begin{enumerate}
	\item for all A$\in$P, $\Xi$(A)$\in$ End(F) is positive definite. This implies that det($\Xi$(A)) $\neq$ 0 for all A$\in$P;
	\item the determinant det($\Xi$(A)) =: N($\Xi$)$^{2}$, N($\Xi$)$\geq$0, does not depend on A. This implies that $\Xi$ is not linear. Indeed, suppose $\Xi$(zA) = z$\Xi$(A) for z$^r\neq$ 1, z$\in$$\C$, then det($\Xi$(A)) = det($\Xi$(zA)) = det(z$\Xi$(A)) = z$^{r}$det($\Xi$(A)), a contradiction.
\end{enumerate}
\textbf{Example}: A general example is provided by maps $\zeta$(A) defined by means of the exponential map on Aut(F)
\begin{eqnarray*}
	e : Lie(Aut(F)) &\rightarrow& Aut(F),\\
	X &\mapsto& e^X,
\end{eqnarray*}
where Lie(Aut(F)) is the Lie algebra of the group of automorphisms on F. \\
Hence, for traceless elements T$^a$, a=1,...,r, of Lie(Aut(F)), and A$\in$P, the operator e$^{iA_aT^a}\in$Aut(F) has determinant one and satisfies all above conditions relative to $\zeta$(A). 
\begin{note} When T$^a$ = 0, so $\zeta$(A) = Id, then the measure $\mu_{m*_{\zeta}g}$ reduces to the Gaussian measure with covariance B$^{-1}_m$ and this means that there is no interaction. When $\zeta$(A) =: $\tau$ does not depend on A, then $\mu_{m*_{\zeta}g}$ reduces to the Gaussian measure with covariance B$^{-1}_{m,\tau}$.
\end{note}
\begin{note} When dim(P) = dim(F) = 1, then e$^{iA}\in\C$ and $\mu_{m*_{\zeta}g}$ reduces to the Gaussian measure with covariance B$^{-1}_{m,e^{iA}}$ = B$^{-1}_m$. This means that our construction is nontrivial for essentially higher dimensional spaces F and P.
\end{note}
Now, it is time to calculate important quantities for probability laws, namely the correlation functions.

\subsection{Correlation functions of interacting probabilities}

Of course, one defines correlation functions and their generating functionals in an analogous manner than usual probability laws. 
\begin{dfn} Let B$^{-1}_m$, B$^{-1}_g$ be the covariances of two Gaussian measures defined on F and P respectively, $\mu_{m*_{\zeta}g}$ be their interacting probability law with interaction $\zeta$, then the \emph{correlation function} of the interacting theory is defined by:
\begin{eqnarray*}
	<f_1...f_N> := \int_F d\mu_{m*_{\zeta}g}(u)\; f_1(u)...f_N(u), \quad f_i \in F^*,\; i=1,...,N,\: N\in \N.
\end{eqnarray*}
\end{dfn}
\textbf{The two-point correlator}: The two-point correlation function is given by:
\begin{eqnarray*}
	<f_1.f_2> &:=& \int_F d\mu_{m*_{\zeta}g}(u)\; f_1(u)f_2(u), \quad f_1,f_2 \in F^*,\\
	&=& N(\zeta)\int_{P} dA\; e^{-\frac{1}{2}B_g(A,A)} \int_{F} du \; e^{-\frac{1}{2}B_m(\zeta(A)u,\zeta(A)u)}\; f_1(u)f_2(u),\\
	&=& \int_P d\mu_g(A) \int_F d\mu_{m,\zeta(A)}(u)\; f_1(u)f_2(u), \\
	&&(\mu_{m,\zeta(A)} \text{ is the Gaussian law with covariance }B^{-1}_{m,\zeta(A)}) \\
	&=& \int_P d\mu_g(A)\,B^{-1}_{m,\zeta(A)}(f_1,f_2), \quad \text{(Wick theorem)}
\end{eqnarray*}
Noticing that 
\begin{eqnarray*}
	B^{-1}_{m,\zeta(A)}(f_1,f_2) = B^{-1}_{m}(f_1\circ\zeta(A)^{-1},f_2\circ\zeta(A)^{-1}), \quad f_1,f_2 \in F^*,
\end{eqnarray*}
it is not difficult to show that the two-point correlator $<f_1,f_2>$ is a Gaussian integral over P with a \emph{nonquadratic} integrand. Its exact calculation seems to be not straightforward but one may use standard perturbative approach by approximating the inverse $\zeta(A)^{-1}$ with polynomials in A.

\section{Conclusion}
We have seen some insights of probability theory in the formulation of QFT within the path integral formalism. This lead us to a probabilistic construction of interacting theories which is obtained by means of conditions compatible to important features of nowadays interacting QFTs such as the path integral description of free theories, and the independency of interaction with rapport to the free theory. The advantage of our construction is that interacting theories are again represented by probability laws and then may be rigourously defined. Our future work will be concerned with further developments of the convolution product construction of interactions, including pertubative calculations of correlations functions of concrete physical system.

\end{document}